# Do ML Experts Discuss Explainability for AI Systems?

## A discussion case in the industry for a domain-specific solution

Juliana Jansen Ferreira
IBM Research
Rio de Janeiro Brazil
jjansen@br.ibm.com

Mateus de Souza Monteiro
IBM Research
Rio de Janeiro Brazil
msmonteiro@ibm.com

## ABSTRACT

The application of Artificial Intelligence (AI) tools in different domains are becoming mandatory for all companies wishing to excel in their industries. One major challenge for a successful application of AI is to combine the machine learning (ML) expertise with the domain knowledge to have the best results applying AI tools. Domain specialists have an understanding of the data and how it can impact their decisions. ML experts have the ability to use AI-based tools dealing with large amounts of data and generating insights for domain experts. But without a deep understanding of the data, ML experts are not able to tune their models to get optimal results for a specific domain. Therefore, domain experts are key users for ML tools and the explainability of those AI tools become an essential feature in that context. There are a lot of efforts to research AI explainability for different contexts, users and goals. In this position paper, we discuss interesting findings about how ML experts can express concerns about AI explainability while defining features of an ML tool to be developed for a specific domain. We analyze data from two brainstorm sessions done to discuss the functionalities of an ML tool to support geoscientists - domain experts - on analyzing seismic data - domain-specific data – with ML resources.

## 1 Introduction

In the digital transformation era, AI technology is mandatory for companies that want to stand out in their industries. To achieve that goal, companies must make the most with domain data, but also combine it with domain expertise. Machine Learning (ML) techniques and methods are resourceful while dealing with a lot of data. But it needs the human input to add meaning and purpose to that data. AI technology must empower users [7]. In the first age of AI, the research aimed to get away from studying human behavior and consider the computer as a tool for solving certain classes of problems [19]. But now, the best results come from the partnership between AI and people where they are coupled very tightly, and the resulting of this partnership presents new ways for the human brain to think and computers to process data. The pairing, or the communication, of machines and people, is the core material for Human-Computer Interaction (HCI) research. Recently, AI research has been recognizing the HCI view on their advances since the human behavior cannot be left out of the context to advance AI research impact on real problems [7][19][29].

The explainability dimension of AI, eXplainable AI (XAI), gains even more importance once people are a component for successful AI application. While researching explainable AI, we observed that different terms are often present in the previous work that sometimes are considered as a synonym of explainable or as a necessary dimension to enable explainability. Interpretability and transparency are constant terms associated with XAI, and they are usually related to algorithms or ML models. Although the keywords help us to search for relevant work in XAI, our goal was to verify if the explanation of AI in the publications has a clear goal not just present any explanation.

AI shows great results dealing with problems that can be cast as classification problems, but they lack the ability to explain their decisions in a way people can understand [21]. Most AI explainability research focuses on algorithmic explainability or transparency [1][7][30][34], aiming to make the algorithms more comprehensive. But this kind of explanation does not work for all people, purpose or context. For those with expertise in ML or maybe only with computer programming, this approach might be enough to build explanations, but not for those people without that technical expertise, such as domain experts.

There is much less XAI research considering usability, practical interpretability, and efficacy on real users [12][34]. The mediation of professionals like designers and HCI practitioners seems even more critical for XAI design [28]. The presence and participation of designers in the early stages of ML models' development presents an interesting approach for XAI. Since designers are the professionals responsible for building the bridge between technology and users, they need to understand their working material. In this case, for XAI, ML models are an essential part of this material for design [17]. HCI presents a lot of methods and approaches that are flexible enough to deal with different design scenarios. The co-design technique is being applied with domain experts [8][32] and also with ML experts or data scientists as users [13][27] to explore explainability functionalities. The explanation challenges are also being tackled in broader aspects that impact the society such as trust (e.g. [[1],[15],[30]]), ethical and legal aspects [16].

It is a challenge to combine ML expertise with domain knowledge to tune ML models for a specific domain. Industries are housing their own AI experts and data scientists [33][35], which is an indicator of the importance of combining AI and domain expertise. There are a set of new roles that AI technology generates, and industries need to adapt and hire AI experts to keep their competitive edge. Some of those new roles created by AI are related to the ability to explain the AI technology in some matter and considering some dimension [14]. One common characteristic of all explanation skills is the contextualization of the AI technology in the business, relate it to the domain. For that, the domain knowledge is the differentiator factor to make general AI solutions tuned for a business needs in the industry.

Our research context is in the oil & gas industry. An essential part of this industry decision-making process relies on experts' prior knowledge and experiences from previous cases and projects. The seismic data is an important data source that experts interpret by searching for visual indicators of relevant geological characteristics in the seismic. It is a very time-consuming process. The application of ML on seismic data aims to augment experts' seismic interpretation abilities by processing large amounts of data and adding meaning to visual features in seismic. The ML tool, in our case, aims to be a sandbox of ML models that can handle seismic data in different ways for different tasks to enable seismic interpretation experts to have meaningful insights during their work.

In this position paper, we discuss some findings about how ML experts can express their concerns about AI explainability while developing an ML tool for supporting the seismic interpretation. We had the opportunity to observe and collect data from two brainstorm sessions where ML developers and ML researchers, some with domain knowledge, discussed features of an ML tool. Although the explainability was not an explicit discussion topic, the concerns about that dimension could be identified in portions of the participants' discourse throughout the sessions.

## 2 Related Work

There are many research efforts regarding explainable artificial intelligence (XAI) in the literature. For this paper, we look over previously published work from different venues (e.g., IUI, CHI, DIS, AAAI, etc.) and databases (e.g., Scopus, Web of Science, Google Scholar) to identify which are the people considered on XAI research. Our research examines two types of people: 1) ML experts, which are people capable of building, training and testing machine learning models with different datasets from different domains, and 2) Non-ML-experts, which are people not skilled with ML concepts that, in some dimension, use ML tools to perform tasks on different domains.

Considering ML experts, there is previous work about supporting the sense-making of the model and data to enable explainability. These studies are often related to delivering explanations through images by showing the relevant pixels (e.g. [22,24]) or regions (e.g. [24]) of pixels from the classifier result. Other works, such as the presented by Hohman et al. [13], uses a visual analytics interactive system, named GAMUT, to support data scientists with model interpretability. Similarly, to Hohman et al. [13], the authors Di Castro and Bertini [11] explore the use of visualization and model interpretability to promote model verification and debugging methods using a visual analytics system. Studies also highlight decision-making before the developing process. One of the applications is to provide support in the process of assertive choosing of the machine learning model. In the work of Wang et al. [27], the authors offer a solution named ATMSeer. Given the dataset, the solution automatically tries different models and allows users to observe and analyze these models through interactive visualization. Lastly, concerning ML experts, but with no visualization, Nguyen, Lease, and Wallace [4] present an approach to provide explanations regarding of annotator mistakes in Mechanical Turkey Tasks.

Concerning non-ML-experts, Kizilcec [30] presented a study on a MOOC platform. The authors show research on how transparency affects trust in a learning system. According to the authors [30], individuals whose expectations (on the grade) were met, did not vary the trust by changing the "amount" of transparency. Besides, individuals whose expectations were violated, trusted the system less, unless the grading algorithm was transparent. Another context-aware example is the work of Smith-Renner, Rua, and Colony [2]. The authors present an explainable threat detection tool. Another work that supports decisions in high-risk, complex operating environments, such as the military, is the work from Clewley et al. [25]. In this context, such use improves the performance of trainees entering high-risk operations [25].

Paudyal et al. [26], on the other hand, present a work in the context of Computer-Aided Language Learning, in which the explanation is used to provide feedback on location, movement, and hand-shape to learners of American Sign Language. Lastly, Escalante et al. [16] explanations happen in the area of human resources, in which routinely decisions are made by human resource departments to evaluate candidates. In ML, this task demands an explanation of the models as a means of identifying and understanding how they relate to decisions suggested and gain insight into undesirable bias [16]. The authors [16] address this scenario by proposing a competition to reduce bias in this ML task.

Works that presents the explanation for non-experts with no context are not unusual. For example, Cheng et al. [15] present a visual analytics system to improve users' trust and comprehension of the model. In another non-context work is from Rotsidis, Theodorou, and Wortham [1], in which the authors show explainability for human-robots interaction. By showing in through virtual reality in real-time, the decision process of the robot is exposed to the user in a debugging functionality. The majority of the ML techniques and tools presented in the literature are designed to support expert users like data scientists and ML practitioners [27] and how visualization has been used widely to explain and visualize algorithms and models (e.g. [13,22,24,27]).

However, the work of Kizilcec [30] shows the complexity in providing explanations or making the algorithm more transparent, especially to non-experts. This fact highlights that the transparency/explainability of models is not static. Instead, it requires a deep understanding of the end-user and the context [32]. Besides, the intelligent system's acceptance and effectiveness depend on its ability to support decisions and actions interpretable by its users and those affected by them [23]. Recent evidence [32] shows that misleading explanation has, consequently, promoted conflicting in reasoning. An explanation design should, therefore, offer the cognitive value to the user and communicate the nature of an explanation relevant to their context [[17],[32]].

Browne [17] presents a reinforcement alternative concerning designing explainability. The author argues that the designers should not only understand the end-user and the context but preferably also participate in the early conceptualization of the ML model. According to Browne [17], with the early participation, the designers benefit from understanding the models more sincerely and allow them to develop early prototyping of ML experiences, i.e., more controllability, testing of the model, and successful explanation strategies.

Towards a user-centered explanation, co-designing the explainable interface appears to be a possible approach to both expert and non-expert end-users. For example, Wang et al. [9] developed a framework using a theory-driven approach. The explanations were focused on physicians with previous knowledge in a decision support system. Similarly, in the same context of Healthcare, Kwon, et al. [8] co-designed a visual analytics system.

Stumpf [32], on the other hand, used co-design to a more abroad intelligent system, a Smart Heating system. In their discovery [32], end-users voted for more explanation through more straightforward and textual explanations. Accordingly, Wang et al. [9] affirm that some explanation structures in specific contexts can be communicated with simpler structures, such as textual explanations or even single lists. On the other hand, some well-structured and complex contexts ask for more elaborate explanations techniques (e.g. [8]), i.e., intelligibility queries about the system state (e.g. [21]) or even inference mechanisms (see [8]) [9]. Other techniques include XAI elements, such as the feature that had a positive or negative influence on an outcome [9].

One work that used co-design for a solution to experts it is the work of Wang et al. [27]. In their work, ML experts participated in the process of elucidating about how they choose machine learning models and what opportunities exist to improve the experience. Another expert-centered work is presented in Hohman et al. [13]. Through an interactive design process with both machine learning researchers and practitioners, the authors emerged a list of capabilities that an explainable machine learning interface should support for Data Scientists.

Finally, Barria-Pineda and Brusilovsky [21] presented the explainability design of a recommender system in an educational scenario. After releasing the system for testing, the authors found that transparency seemed to influence the probability of the student in opening and to attempt the lesson. Other motivations for explainability in the learning context can also be the learning itself (see [26]). Furthermore, the motivation tells a lot about the awareness of the work within the user and the context. Studies that had a perceived context-awareness presented a specific motivation for explaining, that is, choosing appropriate models before developing [27], improving workers' production [3], debugging models [1], training for a military novice [25], among others. Other non-context researches motivated the explanations into generic aspects such as trust (e.g. [[1],[15],[30]]), ethical and legal aspects [16]. Chromik et al. [23], for example, affirm that companies that motivate only through legal compliance will most likely not result in meaningful explanation for users. Legal compliance acknowledges user rights, but it is not enough for users nor our HCI research community [23].

## 3 Our ML tool case

This paper research was designed from the opportunity to observe and hear discussions of a project development team regarding the features for building an ML tool. We observed and collected data from 2 brainstorm sessions where ML developers, ML researchers, and other stakeholders of the ML tool discusses features for that tool. The discussion did not have any orientation to aspects of XAI or any particular feature. They were proposed by the people involved in the project to get a better understanding of the ML tool’s features.

The ML tool project is developed in an industry R&D laboratory and is already being used by oil &gas companies in research projects. We believe it is essential for the research to explain our settings. There was a previous study with some of the participants in the same laboratory where they were invited to reflect and discuss on some ML development challenges, such as XAI [28]. One of the authors of this paper participated in this previous study as an HCI researcher and saw the discussions of the ML tool an opportunity to reflect and discuss ML challenges in a real project context. Therefore, she participated in the session as an observer without any intervention or mediation and collected the data used to discuss in this paper.

### 3.1 Research Domain Context

The ML tool of our case aims to aid seismic interpretation, which is a central process in the oil and gas exploration industry. This practice main goal is supporting decision-making processes by reducing uncertainty. To achieve that goal, different people work alone and engage in multiple informal interactions and formal collaboration sessions, embedding biases, decisions, and reputation. Seismic interpretation is the process of inferring the geology of a region at some depth from the processed seismic data survey[1] . Figure 1 shows an example of seismic data lines (or slices), which is a portion of seismic data, with an interpretation about visual indicators of a possible geological structure called salt diapir.

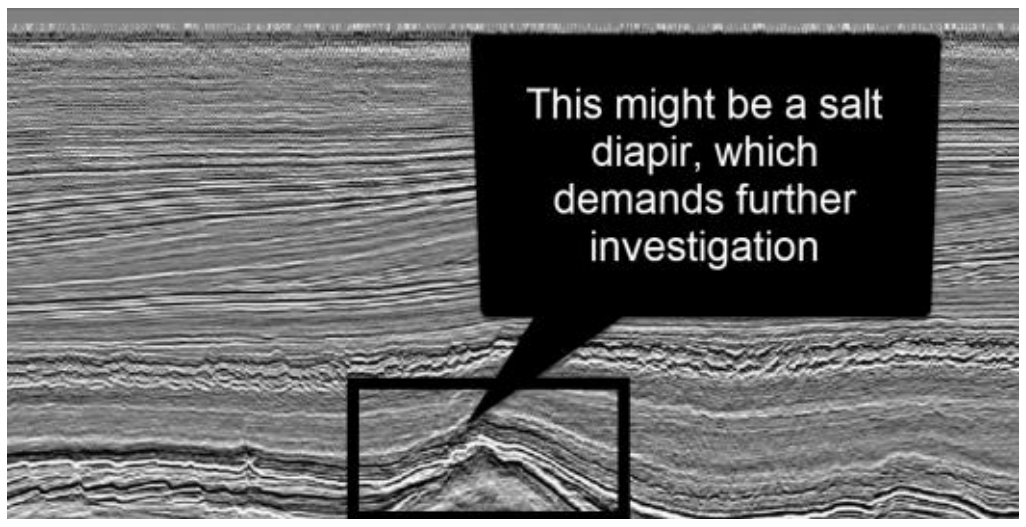

**Figure 1: Seismic image example (Netherland – Central Graben – inline 474)**

In the same industry R&D laboratory, ML experts and researchers are exploring the possibilities of combining ML for exploring seismic data. It is important to say that seismic data are mainly examined visually. It commonly has other data to compose the seismic interpretation, but the domain expert analyzes, interprets the seismic imagens to identify significant geological characteristics. Therefore, there is research focusing on image analysis aspects rather than geophysical or geological discussions. [5][6]. Plus, there is research on exploring additional texture features that are prominent in other domains but have not received attention in the seismic domain yet. Namely, they investigated the ability of Gabor Filters and LBP (Local Binary Patterns) – this last, widely used for face recognition – to retrieve similar regions of seismic data [6]. Still exploring the visual aspects of seismic data, there is research on generating synthetic seismic data from sketches [31] and on using ML to improve the seismic image resolution [10].

## 3.2 About the ML professionals

In total, there were eleven (11) ML professionals as participants on the ML tool discussions: ML Developers (7) that were involved in the ML tools' discussions and where directly involved in its development. ML Researchers (2) that were involved in the discussion about the ML tool, but not directly involved in the development, Domain Expert (1) that is a member of the technical team (not expert from the industry), but with deep understanding of the domain data and domain practice with that data, and a facilitator (1) that facilitate the brainstorm session without influencing on the discussion content.

As aforementioned, for this research, we have four (4) participants that already collaborated in a previous study [28]. Three (3) of them have more than seven years of experience with ML development and research, and they have been working in the oil & gas industry for more than one year (1 of them for more than four years). Those participants have been working with the domain data in question (seismic data) for a while and have been exploring different aspects of it with ML technology [5][6][10][31]. The other participants are also experienced ML developers or experts having at least three years of experience in the industry, plus academic experience.

## 3.3 Brainstorm sessions

The data we collected for this paper analysis was produced during two brainstorm sessions for the development of a domain-specific ML tool. With participants' consent with the data collection before the sessions, and they were aware that it was going to be used for research publication.

The ML tool under development is an asset from a larger project with industry clients; therefore, its development aims to support real domain practices. The brainstorm sessions were organized by the ML tool's development team from the laboratory. It was not scheduled to produce data for our study in particular but is presented as an enriching opportunity to investigate if and how ML Experts discuss AI explainability in while they are building an ML tool.

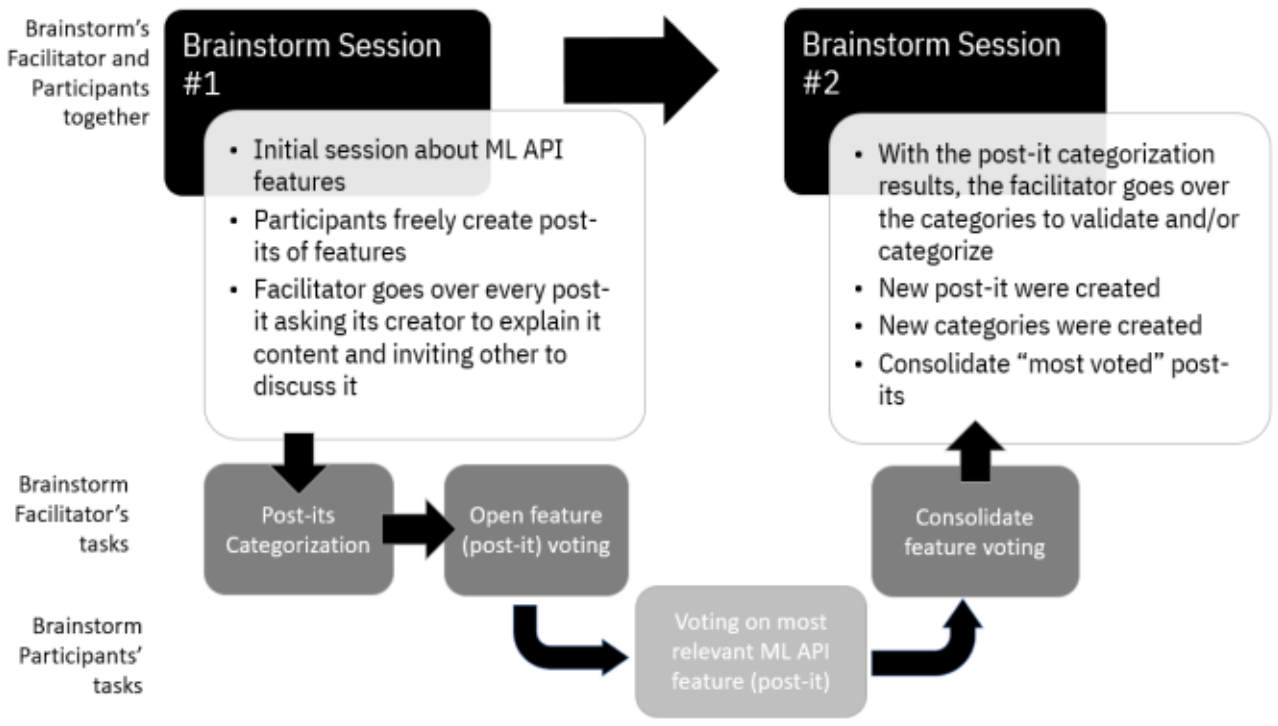

**Figure 2. Brainstorm Sessions plan**

There were two brainstorm sessions organized to discuss the ML tool's features. The facilitator organized activities to support individual inputs and collaborative discussions (Figure 2). Between the sessions, there was a voting activity to prioritize the discussion for the second session. The sessions were performed in an online collaboration tool[2]. The content of the collaboration tool was discarded as study data because one participant modified it without the facilitator orientation. Therefore, this study data was the videos of the session. Some of the participants were not physically present, participating through a videoconferencing system and the online tool.

[1] https://www.britannica.com/science/seismic-survey

[2] https://mural.co/

## 4. Data Analysis

As aforementioned, we used the sessions' videos as our study data. We transcribed the audios from both videos (session 1: 2h and session 2: 1.5h, respectively) and tagged the quotes of every participant of the sessions. We wanted to identify the XAI aspects of the discourse and relate it to the participant who brought it to the discussion. We considered the data from both sessions as one dataset because we wanted to analyze the discourse of participants throughout the discussion about the ML tool's features.

For the data analysis, we used a qualitative approach since we are still framing concerns about XAI on ML tools' development. Our goal was to identify the critical ideas that repeatedly arise during the ML professionals' discussion of an ML tool's features. We used the discourse analysis method that considers the written or spoken language concerning its social context [18] (pp. 221, [20]). We did start by doing some content analysis (pp. 301, [20]) to verify the frequency of terms, cooccurrences, and other structural markers. But since the topic of the discussion was broader – ML tool's features – this did not provide relevant findings. Therefore, we changed to discourse analysis, which goes beyond looking at discussions of words and contents to examine the structure of the conversation, in search of cues that might provide further understanding (pp. 221, [20]).

## 4 Discussion about AI Explainability

We started our data analysis trying to tag the participants' quotes with the codes "aid-XAI" or "harm-XAI" (aid or harm eXplainable AI). Then, we notice that any categorization of the data we had was not possible without further feedback from the person who said the quote. Therefore, we decide to tag the quotes that had in the discourse features or concerns related to AI explainability. We selected a total of 25 quotes from approximately 3.5h of audio transcriptions. Considering that the brainstorm session had a broader goal of discussing the ML tool's features, we believe those quotes point to an exciting direction for our research to investigate "Do ML Experts Discuss Explainability for AI Systems?". The discussion did not have any intervention or bias towards explainability concerns, which allow us to see if and how AI explainability would be included in their development discussion.

From the 25 quotes, 13 were from those three ML professionals that have more experience with ML development and also experience working with the domain data (seismic data). We learned that professionals that have ML+Domain knowledge combined might be more capable of having an overall vision of how the AI system will impact the domain and its experts. The quotes indicate concerns about XAI without any mention of the specific topic. The theme was of genuine concern from those professionals, and it was present in their discourse while developing an AI system for geoscientists. In this position paper, we selected a few of those quotes to discuss the concerns ML developers are expressing about AI explainability while thinking about features for an ML tool.

The discussion for the ML tool was sometimes conflicting about who was the user (or user) for that ML tool. In the quote below, one participant was considering two users: an ML expert and a data scientist. In his discourse, it is aligned with previous research about ML models' interpretability [11][13] and understanding the data that ML models handle [22,24]. The visualization of trained model and the visualization of the data with its metrics could be a way to explain an XAI scenario for ML experts and data scientists. This kind of feature could be a pointer to further discussions on XAI:

> [...] a visualization, feature "I'm a machine learning guy and I want to see the trained model"; "I'm the data guy and I want to see the data [...] I want to correctly visualize the data [...] how is this data spatially distributed [...] visualize the metrics. [...].

In the next quote, a participant comment on a new trend in oil & gas companies of training geoscientists on machine learning. This trend aims to combine the ML tools potential to handle a lot of data and the domain expert tacit knowledge and experience to tune the pair model-data to have the best results with ML. Not only quantitative results (best ML model accuracy) but qualitative results when that domain expert with ML learning knowledge can make the best of model-data by understanding the meaning of the results. There are new roles of "explainers" in AI [34] that will make the technology fit the domain in which it is applied. By having the understanding model and domain data, they are equipped to define the necessary explanations in a domain:

> [...] what happens in these companies now is that they are hiring geophysicists and giving a machine learning course, and I also think the same guy may be acting depending on the role he's playing at that time [...].

The understanding of the algorithms and the ML workflows has been the focus of most XAI research [1][7][31][35]. The trails on what data goes into which model and which was the output result can support the decision about how to fit the model and data were for a particular case. In the next quote, a participant places a concern about the timeline and resolution of the seismic data. Those are parameters of the seismic data that could help the building a better ML tool. A comparison feature could be considered a way to explain what is available, what was in fact, used by the ML tool and why:

> [...] you have to imagine that you have seismic data from 20 years ago, as usual, and you have a new seismic data that has a different resolution [...] For you to be able to compare things, you need to have a grid there and start comparing things. All the information that goes in there needs to be useful [...]

The participants were mostly ML developers; therefore, they are used to handle ML models and data like one type of user considered for the ML tool under development. The quote above shows a participant finding a solution to their users the same as him, as the user thinks as a good solution. This seems an interesting approach: to use existing tools that somehow explain the ML results and see if it works for other users. Combining this initial input with co-designing approaches [13][27], the investigation of what works as an explanation for every user could present promising research results:

> [...] something like Jupyter does. You have a report that says, "For this data here I had this result," the views and the guy can follow more or less [...]

## 5 Final Remarks and future work

In this position paper, we aim to use the data collected from a real ML tool's development project brainstorm to discuss if and how ML experts express concerns about AI explainability while defining features of an ML tool to be developed. It was not a controlled study with users. We analyze data from two brainstorm sessions done to discuss the functionalities of an ML tool to support geoscientists - domain experts - on analyzing seismic data - domain-specific data – with ML resources. It was serendipity that one of the authors got aware of the discussion and that all participants agree that she could be present and collect the data for this research.

The data collected was tough to transcript because the brainstorm sessions were used to structure all the participants understanding the ML tool, user, and features. Therefore, sometimes participants did not make complete sentences, or the sentences were incomprehensible. As mentioned in the Data Analysis session of this paper, we started the data analysis with content analysis [20] but changed to discourse analysis [18] to analyze the data. But while analyzing word frequency, we generate the word cloud presented in Figure 3. The most frequent word was "you" which was used by participants to present their ideas.

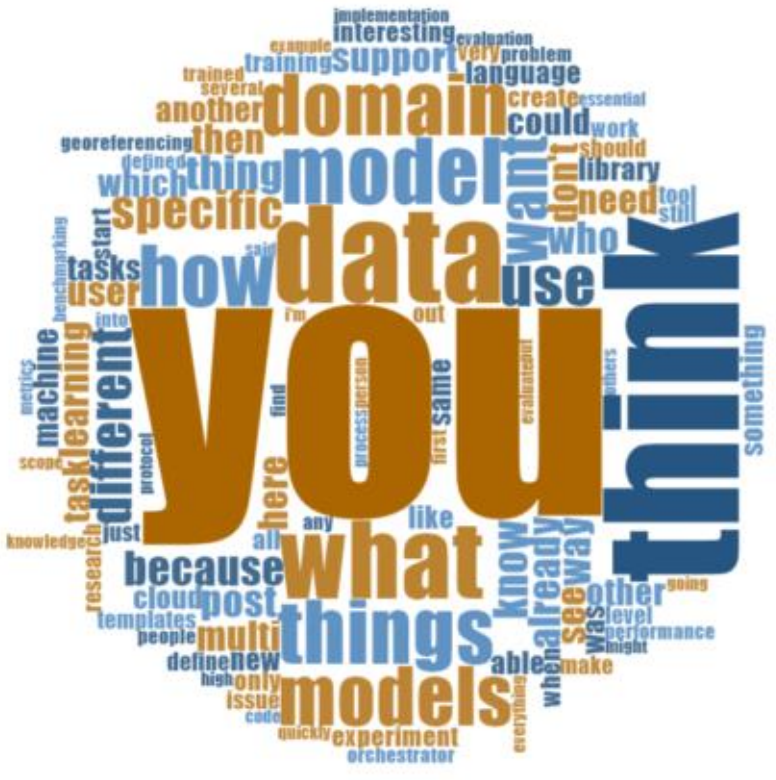

**Figure 3. Word cloud from transcripts**

Considering that ML professionals were one of the potential users for the ML tool, it is interesting that ML developers did not use the first person in their phrases, but the third person " you". An investigation path was to check with those ML professionals if they thought of themselves as a possible user to the ML tool and how it would affect the discussion about its features. Using design techniques, such as co-design [13][27], to explore those scenarios with ML professionals as users could open different discussions topics. Maybe concerns about explainability would appear more once developers are in users' place.

In a previous study in the same R&D lab, mediation challenges were identified for the development of deep learning model [28]. One exciting aspect of that earlier study was that once the ML professional considered his ML solution in a real context, new concerns about the impact on people and explanations were identified. In this study, the ML professionals have a real context where their ML tool will be applied, but we believe they are still very distant from the consequences the ML tool might have on the user decision-making. The study reported in [28], the context and its impacts were easier to relate (ML to support hand-written voting process using MNIST dataset). For the oil & gas domain, for example, the effect of a wrong decision cannot be so easily foreseen. This could be an approach for investigating the mediation challenges [28].

Explanations are social, and they are a transfer of knowledge, presented as part of a conversation or interaction, and are thus shown relative to the explainer's (explanation producer) beliefs about the 'explainee's' (explanation consumer) beliefs. [34]. XAI needs social mediation from technology builders to technology users and their practice [28]. We believe the explanation cannot be generic. The design of a "good" explanation needs to take into account: who is receiving the explanation, what for and in which context the explanation was requested.

This initial study opened paths to many exciting kinds of research, not only associated with XAI. For the XAI research, as future work, we intend to investigate AI explanations considering those three dimensions (who + why + context). The investigation of XAI considering those dimensions shows promising paths for designing AI systems considering different scenarios. Industries are training their domain experts on ML tools, but what about capacitate ML experts on data and domain practice before building ML solutions? It might enable the ML expert to design the solution aware of how it will impact the domain and the people involved.

Other promising research path is to address the XAI topic explicitly with ML professionals as part of the design material for developing AI systems. The mediation challenges identified by Brandão et. al [28] are an initial pointer for that XAI discussion with ML professionals . As our first study, we plan to go back to the same group participants and discuss AI explainability to verify what kind of feature and concerns are raised once we point to the specific topic.